          [1999/12/02 1.01 (PWD)]
\documentclass[a4paper,twocolumn]{esapub}
\usepackage{graphics}

\usepackage{natbib}
\newcommand{\be}{\begin{equation}}
\newcommand{\ee}{\end{equation}}
\title{Unidentified Gamma-Ray Sources and Microquasars}

\author{G.E. Romero}
\affil{\it IAR, C.C. 5, 1894
Villa Elisa, Argentina. E-mail: romero@venus.fisica.unlp.edu.ar}
\author{I.A. Grenier}
\author{M.M. Kaufman Bernad\'o}
\author{I.F. Mirabel}
\affil{\it CEA/DSM/DAPNIA/Service d'Astrophysique, Centre
d'Etudes de Saclay, F-91191 Gif-sur-Yvette, France}
\author{D.F. Torres}
\affil{\it Lawrence
Livermore National Laboratory, 7000 East Ave., L-413, Livermore,
CA 94550, USA}

\begin{document}


\maketitle

\begin{abstract}
Some phenomenological properties of the unidentified EGRET detections suggest that there are two distinct groups of galactic gamma-ray sources that might be associated with compact objects endowed with relativistic jets. We discuss different models for gamma-ray production in both microquasars with low- and high-mass stellar companions. We conclude that the parent population of low-latitude and halo variable sources might be formed by yet undetected microquasars and microblazars. 
\end{abstract}

\section{Introduction}

Population studies indicate that there are at least three different groups of galactic gamma-ray sources (Gehrels et al. 2000; Grenier 2001, 2004). One is a group of bright sources distributed along the galactic plane, with a concentration toward the inner spiral arms. These sources are well-correlated with young stellar objects and star-forming regions (Romero et al. 1999, Romero 2001). We will call these sources Gamma-Ray Population I (GRP I). There is another group of sources distributed around the Galactic Center, forming a halo with a radius of $\sim 60^{\circ}$. These sources are softer and more variable than GRP I sources. We will call this group Gamma-Ray Population II (GRP II). Finally, there is a group of sources correlated with the Gould Belt (Grenier 1995, 2000). These are nearby, relatively young and weak sources. We will name them the Local Gamma-Ray Population (LGRP). 

Recent variability studies (Nolan et al. 2003; Torres et al. 2001a,b) show that GRP II sources are very variable. There is also a subgroup of variable sources among those in the GRP I. This variability, with timescales from weeks to months, indicates that the counterparts are compact or small objects. Supernova remnants, OB star forming regions, and other extended sources are ruled out. Isolated pulsars, which are known to be steady sources, are also discarded. 

Recently, Kaufman Bernad\'o et al. (2002) have suggested that the parent population of GRP I sources are microquasars with high-mass stellar companions and jets that form a small angle with the line of sight, i.e. the ``microblazar" population suggested by Mirabel \& Rodriguez (1999). In the Kaufman Bernad\'o et al. (2002) model, the gamma ray emission is produced by inverse Compton (IC) up-scattering of UV photons from the companion stars, as suggested by Paredes et al. (2000) for the microquasar LS 5039 (see also Bosch-Ramon \& Paredes 2004). More recent models (Romero et al. 2003) incorporate the effect of hadrons in the interaction of the jet with the wind of the star. In the present paper, we extend the Kaufman Bernad\'o et al. (2002) proposal to encompass both GRP I and GRP II variable sources. We suggest that old low-mass microquasars, either ejected from the galactic plane or from globular clusters, might be the counterparts of the GRP II. For GRP I sources, we suggest that a subgroup of them (the most variable ones) might be high-mass microquasars. Some of them might have soft TeV counterparts. In the next sections we present some models for these objects.    

\section{Unidentified EGRET sources: relevant properties}

GRP I sources in the Third EGRET Catalog (Hartman et al.~1999) have an average photon index $\Gamma=2.18\pm0.04$. Their $\log N-\log S$ plot is a steep power law typical of young objects concentrated in the inner spiral arms. The expected (isotropic) luminosity range for these sources goes from $10^{34}$ to $10^{36}$ erg s$^{-1}$. There are 17 variable sources in this group. 

\begin{figure}[t]
\centering
\begin{flushleft}
\resizebox{8cm}{!} {\includegraphics{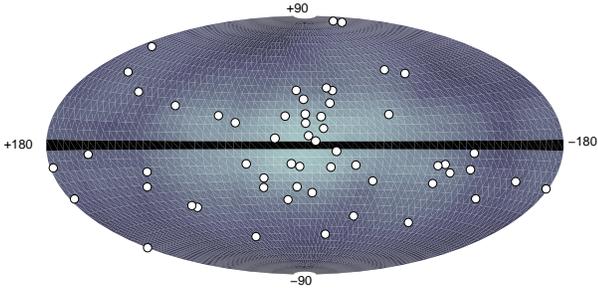}}
\caption{Halo sources in the 3EG catalog. From Grenier (2004). \label{fig1}}
\end{flushleft}
\end{figure}

GRP II sources are softer ($\left\langle \Gamma\right\rangle\sim 2.5$) and significantly more variable. Actually, they are even more variable than the isotropic population of extragalactic sources (Nolan et al. 2003). There are $45\pm 9$ sources in this group. They have a distribution consistent with a spherical halo around the Galactic Center with a density profile that is nearly flat up to a radius $R\sim2.5$ kpc and then falls as $R^{-3.5}$. Their inferred 1-sr luminosities are in the range $10^{33}-10^{36}$ erg s$^{-1}$. The distribution of these sources is shown in Fig. \ref{fig1}.

GRP I sources are young, with ages of a few Myr. On the contrary, GRP II should have ages from hundreds of Myr to Gyr.    

\section{Gamma-ray emission from microquasars: massive companions}

It is usually said that microquasars, i.e. X-ray binaries with relativistic jets, are scaled-down versions of extragalactic quasars. In fact, there are many similarities, especially regarding the nature of the accretion disk/jet coupling. However, with respect to the high-energy emission, not the similarities but the differences between these two types of objects are most interesting. The main difference is of course the presence of a star companion in the case of microquasars. This star provides a gravitational field that can exert a torque on the accretion disk of the compact object that can result in the precession of the jets, and hence in the periodic variability of the non-thermal emission (Kaufman Bernad\'o et al. 2002). In addition to the gravitational field, a massive star has a strong photon field (peaked in UV in the case of an early O star) and a matter field in the form of a strong particle wind. The relativistic jets ejected from the surroundings of the compact object must traverse these fields, so interactions are unavoidable. 

\begin{figure}[t]
\centering
\begin{flushleft}
\resizebox{8cm}{!} {\includegraphics{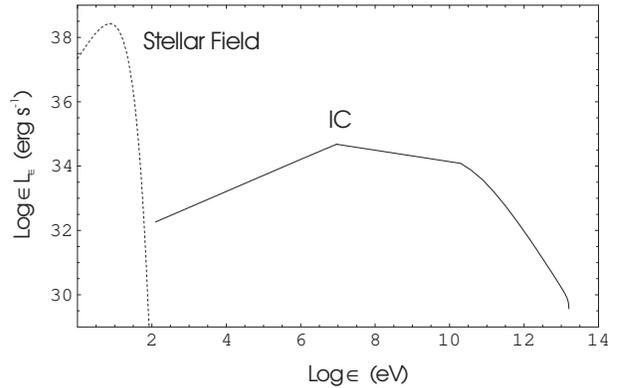}}
\caption{Inverse-Compton spectral energy distribution for a microquasar with a massive stellar companion. Leptons in the jet are assumed to have a power-law energy distribution with an index $\alpha=2$ and a high-energy cut-off at multi-TeV energies. Notice the softening of the spectrum at high-energies due to the Klein-Nishina effect. \label{fig2}}
\end{flushleft}
\end{figure}

IC up-scattering of UV stellar photons produce a high-energy emission that depends on the particle energy distribution, geometry of the jet, orbital parameters, and the spectral type of the companion star (Kaufman Bernad\'o et al. 2002, Georganopoulos et al. 2002, Romero et al. 2002). In Fig. \ref{fig2} we show results of calculations in the case of a microquasar with an O9 I star, similar to the companion star in Cygnus X-1. We have considered the interaction of the photon field with an homogeneous jet with a power-law particle spectrum of index $\alpha=2$ and total power proportional to the accreting rate $P_{\rm jet}=q_{\rm j} \dot{M} c^2$, where  
$q_{\rm j}\sim 10^{-3}$ is the disk/jet coupling parameter (Falcke \& Biermann 1995). The remaining parameters are as in Romero et al. (2002), except for the high-energy cut-off of the particles, which is here in the TeV range, corresponding to electron Lorentz factors of $\gamma=10^{7.5}$. The existence of TeV electrons in microquasar jets have been recently inferred from the detection of synchrotron X-ray emission in XTE J1550-564 (Corbel et al. 2002).     

The spectral energy distribution of the IC emission has a peak at MeV energies, due to the effect of losses on the particle spectrum. At energies above $10^{10}$ eV the Klein-Nishina effect becomes important and the spectrum softens significantly. In the EGRET energy range, the luminosity (per sr) inferred in the observer system is between $10^{34}$ and $10^{35}$ erg s$^{-1}$, for a Doppler boosting $D\sim 1$, quite consistent with the expected values for low-latitude unidentified EGRET sources. The photon spectral index in this range is $\Gamma\sim 2$. At TeV energies, where the spectral index is $\Gamma>3$, the luminosity is $L(E>1\;{\rm TeV})\sim 10^{32}$ erg s$^{-1}$. Hence, one of these microquasars located at a few kpc might be a detectable TeV gamma-ray source. 

\begin{figure}[t]
\centering
\begin{flushleft}
\resizebox{8cm}{!} {\includegraphics{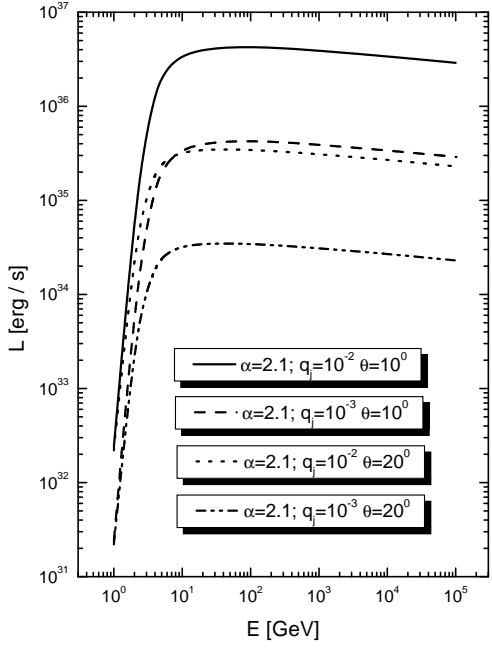}}
\caption{Spectral energy distribution produced by the interaction of a hadronic jet of a microquasar with the wind of the massive stellar companion. Curves for different jet/disk coupling constants and viewing angles are presented. The index in the proton power-law distribution is 2.1 in all cases. \label{fig3}}
\end{flushleft}
\end{figure}

If there are relativistic protons in the jets of some high-mass microquasars, $pp$ interactions with particles in the stellar wind can result in significant gamma-ray emission. In order to make quantitative estimates of the $\pi^0$-decay gamma-ray luminosity we have considered an O star with a mass loss rate of $10^{-5}$ $M_{\odot}$ yr$^{-1}$ and a terminal wind velocity $v_{\infty}=2500$ km s$^{-1}$. The radial dependence of the wind velocity is $v(r)=v_{\infty}[1-(r_*/r )]$, where $r_*=35$ $r_{\odot}$. The jet is assumed to be perpendicular to the orbital plane, with a bulk Lorentz factor of 5, and injected at 50 gravitational radii ($r_{\rm g}$) above the black hole (mass $M_{\bullet}=10$ $M_{\odot}$), with an initial radius of 5 $r_{\rm g}$. The jet is allowed to expand laterally, conserving the number of particles. Protons in the jet have a power-law spectrum of index $\alpha$ in the co-moving frame, with a high-energy cutt-off above 100 TeV. The orbital radius is 2 $r_*$. We have calculated the gamma-ray emission for $\alpha=2.1$ and different combination of viewing angles and disk/jet coupling constant (the accretion rate is the same as in the previous example: $\dot{M}\sim 10^{-8} M_{\odot}$ yr$^{-1}$). The results are shown in Fig. \ref{fig3}.  

We see that hadronic high-mass microquasars can be not only strong sources in EGRET energy band, but also strong TeV sources. Contrary to the case of IC TeV microquasars, they should be sources with a hard spectrum that would mimic the spectrum of the parent population of protons. This prediction can be used to differentiate them. 
  
Gamma-ray variability naturally arises in high-mass microquasar models through jet precession, shocks, and variable accretion rate onto the compact object. 

High-mass microquasars are originated in star forming regions, hence a concentration toward the inner spiral arms is expected. Their proper motions can present velocities of a few hundred kilometers per second, but because of their rapid evolution and short lifetime they are not expected to be found far beyond the galactic disk (Rib\'o et al. 2002).  

\section{Gamma-ray emission from microquasars: halo sources}

Low-mass microquasars are not constrained to remain in the galactic plane. They can be ejected by kicks imparted in the natal supernova explosions to galactocentric orbits (Mirabel \& Rodrigues 2003). Alternatively, some low-mass microquasars might be born in globular clusters from where they might be injected in large galactic orbits (Mirabel et al. 2001).

\begin{figure}[t]
\centering
\begin{flushleft}
\resizebox{8cm}{!} {\includegraphics{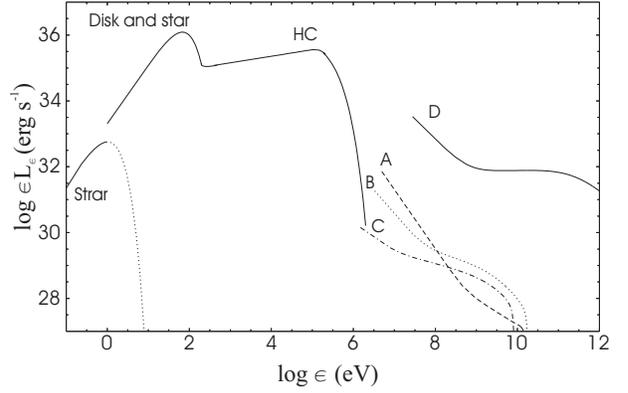}}
\caption{Inverse-Compton spectral energy distribution for a microquasar with a low-mass stellar companion. Leptons in the jet are assumed to have a power-law energy distribution with an index $\alpha=2.3$ and a high-energy cut-off at $\gamma_{\rm max}=10^4$. Different curves correspond to different viewing angles. A: $5^{\circ}$, B: $15^{\circ}$, C: $30^{\circ}$, and D: $1^{\circ}$. This latter case is a microblazar with $\alpha=1.6$, $\gamma_{\rm max}=10^{7}$, and a bulk Lorentz factor of 10 instead of 3 as in the other models.  The disk/jet coupling parameter is $q_{\rm j}=0.01$ in all cases. Effects of the interaction with photons from the accretion disk and a hot corona around the black hole are considered. \label{fig4}}
\end{flushleft}
\end{figure}

IC up-scattering of external photons fields in the case of low-mass microquasars will not be as efficient as in the case of the high-mass objects, mainly due to the reduced stellar luminosities. We have performed calculations of the high-energy gamma-ray spectrum of low-mass microquasars within a variety of models. In Fig.\ref{fig4} we show some results. We have considered here not only the stellar photon field (a K7 V-M0 V star, similar to the donor star in XTE J1118+480) but also the accretion disk field and a hard X-ray component (HC, in the figure), with a power-law spectrum and an exponential cut-off that represents a corona or ADAF region around the central object. The jet is homogeneous, with a bulk Lorentz factor of 3, and the disk/jet coupling parameter is $q_{\rm j}=0.01$. The electron energy distribution is a power-law with index $\alpha=2.3$ and a high-energy cut-off at $\gamma_{\rm max}=10^4$. In the calculations we used the full Klein-Nishina cross section and we took into account the effect of losses into the particle spectrum. The resulting high-energy tail is shown for three different viewing angles: $5^{\circ}$, $15^{\circ}$, and $30^{\circ}$. The spectra at EGRET energies are soft, but the luminosities (per sr) are smaller than what is inferred for GRP II sources.   


Looking for higher luminosities, we have considered the case of a microblazar with a viewing angle of only $1^{\circ}$ and a Lorentz factor of $10$, an electron energy distribution with index $\alpha=1.6$, and a high-energy cut-off $\gamma_{\rm max}=10^7$. The rest of the parameters remain with the same value as in the previous models. The calulated spectrum is shown as curve D in Fig. \ref{fig4}. We see that in this case we get luminosities of around $10^{33}$ erg s$^{-1}$ (per sr) at EGRET energies. If the emission is highly beamed ($<0.1$ sr), luminosities similar to those of GRP II sources are obtained. The addition of a magnetic field does not significantly affect the results at GeV energies unless $B>100$ G (Bosch-Ramon, Romero \& Paredes 2004). With high magnetic fields, by the other hand, strong radio sources are produced, which is at odds with the absence of radio counterparts. More detailed models will be presented elsewhere (Kaufman Bernad\'o et al., in preparation) but the results seem to suggest that the population of gamma-ray sources in the halo might be formed by low-mass microblazars, i.e. sources with highly relativistic and well-collimated jets forming small viewing angles.


\section{Conclusions}

There are variable gamma-ray sources in the Galaxy. There are reasons to think that at least two different populations of such sources exist: one population of young sources in the galactic plane, and one population of old sources at high latitudes, forming a halo around the galactic center. We suggest that these two groups of sources correspond to high-mass microquasars and low-mass microblazars, respectively. Our calculations indicate that the required high-energy spectra and luminosities can be produced by these objects. In a separate communication (Kaufman Bernad\'o et al. in preparation), a detailed discussion of the source statistics will be presented. Some of these sources, especially those at low latitudes with a peak at MeV energies (see Fig. \ref{fig2}), might be detectable by INTEGRAL. The recent findings by Combi et al. (2004) seem to be a promising step in this direction.

\subsection*{Acknowledgments}

This work has been supported by CONICET (PICT 0438/98)
and Fundaci\'on Antorchas. GER thanks the organizers for partial support to facilitate his attendance to the meeting and the SECyT (Argentina) for a Houssay Prize.


\end{document}